\documentclass[letterpaper]{article} 
\usepackage{aaai25}  
\usepackage{times}  
\usepackage{helvet}  
\usepackage{courier}  
\usepackage[hyphens]{url}  
\usepackage{graphicx} 
\usepackage{natbib}  
\usepackage{caption} 
\usepackage{algorithm}
\usepackage{algpseudocode}

\usepackage{amsmath}
\usepackage{amssymb}
\usepackage{mathtools}  
\usepackage{booktabs}       
\usepackage{multibib}

\usepackage{newfloat}
\usepackage{listings}
\DeclareCaptionStyle{ruled}{labelfont=normalfont,labelsep=colon,strut=off} 
\newcommand{\mname}{AudioGenX}

\title{\mname: Explainability on Text-to-Audio Generative Models}
\author {
    Hyunju Kang\textsuperscript{\rm 1}\thanks{Equal contribution.},
    Geonhee Han\textsuperscript{\rm 1}\footnotemark[1],
    Yoonjae Jeong\textsuperscript{\rm 2},
    Hogun Park\textsuperscript{\rm 1}\thanks{Corresponding author.}
}
\affiliations {
    \textsuperscript{\rm 1}Department of Artificial Intelligence, 
Sungkyunkwan University, Suwon, Republic of Korea\\
    \textsuperscript{\rm 2}Audio AI Lab, NCSOFT, Seongnam, Republic of Korea\\
    \{neutor, gunhee8178\}@skku.edu, hybris75@gmail.com, hogunpark@skku.edu}

\begin{document}
\maketitle

\begin{abstract}

Text-to-audio generation models (TAG) have achieved significant advances in generating audio conditioned on text descriptions. However, a critical challenge lies in the lack of transparency regarding how each textual input impacts the generated audio. To address this issue, we introduce \mname{}, an Explainable AI (XAI) method that provides explanations for text-to-audio generation models by highlighting the importance of input tokens. \mname{} optimizes an $Explainer$ by leveraging factual and counterfactual objective functions to provide faithful explanations at the audio token level. This method offers a detailed and comprehensive understanding of the relationship between text inputs and audio outputs, enhancing both the explainability and trustworthiness of TAG models. Extensive experiments demonstrate the effectiveness of \mname{} in producing faithful explanations, benchmarked against existing methods using novel evaluation metrics specifically designed for audio generation tasks.

\end{abstract}
\section{Introduction}

Text-to-audio generation models (TAG)~\cite{kreuk2022audiogen, ziv2024masked, yang2023diffsound, liu2023audioldm,  schneider2023mo} have emerged as a pivotal technology in generative AI, enabling textual content to be transformed into an auditory experience. Although models such as AudioGen~\cite{kreuk2022audiogen} excel at generating high-quality audio based on textual prompts, a critical challenge remains: the lack of transparency in how each textual input affects the generated audio. Consequently, users may struggle to trust the model, making it essential to provide explanations for the TAG task. Explainability provides several key advantages. First, it enhances awareness of how input tokens affect the model’s outputs, enabling users to ensure that the model emphasizes the correct aspects of the text. Second, it provides actionable insights to support the decision-making about which elements to modify and to what extent in the audio editing process. Third, analyzing generated explanations can aid with debugging and identifying potential biases. Accordingly, this study argues that the ability to quantify the importance of textual inputs in TAG models is crucial to being able to unambiguously assess and communicate their value.

\begin{figure}
    \center
    \includegraphics[width=\columnwidth] {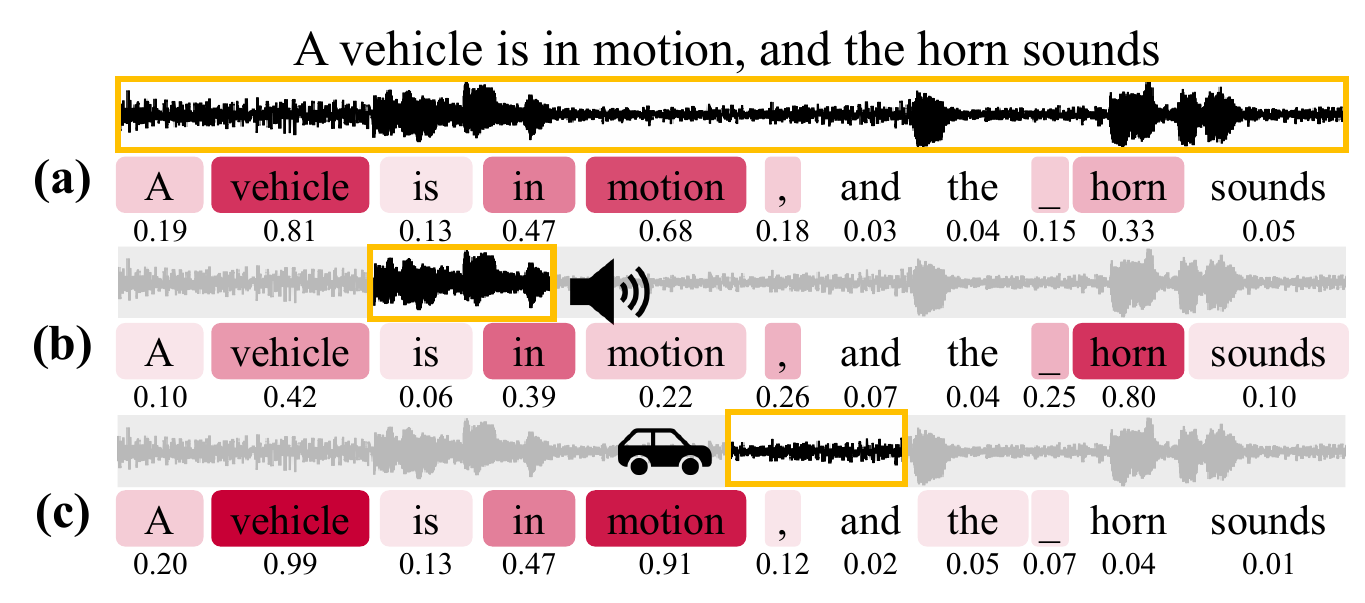}
    \caption{A comprehensive explanation provided by \mname{} for the entire audio in (a). Granular explanations for the interval from 1 to 1.5 seconds in (b) and from 2.5 to 3 seconds in (c), respectively.}
    \label{intro}
\end{figure}

While approaches specifically tailored for explaining TAG models are limited, recent research has explored methodologies for calculating the importance of input tokens in large-scale transformer-based models. Cross-attention layers in multi-modal architectures, such as those in TAG models, are widely regarded as critical for integrating textual and auditory information, while also enhancing explainability by revealing how information from one modality influences another. A notable method~\cite{abnar2020quantifying} utilizes attention weights and aggregates them across all layers to approximate the importance of each input token. However, attention scores alone are not considered reliable for causal insights, as they do not directly indicate how perturbation to specific inputs influences the output. Recently, AtMan~\cite{deiseroth2023atman} introduced a perturbation method that suppresses the attention score of one token at a time to observe the impact of each input on output prediction. This single-token perturbation approach, however, may overlook interactions between multiple tokens. Consequently, it provides less reliable explanations in scenarios where the model heavily relies on the contextual relationships between multiple tokens, leading to an oversimplification of the model’s behavior.

To address the challenge of faithful explanations, causal inference theory, encompassing factual and counterfactual reasoning, is often utilized~\cite{pearl2009causal}. These two approaches aim to identify impactful input information in different ways. Factual reasoning focuses on identifying critical input information that reproduces the original prediction, whereas counterfactual reasoning~\cite{tan2022learning, ali2023explainable, kenny2021post} seeks to determine crucial input information that, if absent, would change the prediction. Given their differing assumptions, these reasoning approaches can be employed together as complementary frameworks to generate more faithful explanations. However, prior research has yet to investigate the feasibility of applying factual and counterfactual reasoning within TAG models.

To provide faithful explanations for the TAG model, we introduce \mname{}, a perturbation-based explainability method leveraging factual and counterfactual reasoning. Our approach utilizes the latent representation vectors in TAG models to observe the effects of factual and counterfactual perturbations. These perturbations are applied in the cross-attention layer using a soft mask, enabling the simultaneous perturbation of multiple tokens' attention scores. More importantly, the mask itself serves as an explanation, with its values quantifying the importance of the textual input. We optimize the mask through a gradient descent method guided by our proposed factual and counterfactual objective functions. To mitigate the high computational cost of calculating gradients for the entire sequential audio, we enhance efficiency by decomposing the explanation target into individual audio tokens. This approach enables us to customize the explanation range of generated audio interactively, providing comprehensive explanations for the entire audio or more granular explanations for specific segments of interest, depending on user demand. For instance, in Figure~\ref{intro}, (a) provides a comprehensive explanation for the entire audio, indicating a strong relation to vehicle motion. By focusing on a specific interval in (b) and (c), \mname{} captures the different contexts of each audio segment and delivers contextually accurate explanations accordingly. Extensive experiments demonstrate the faithfulness of our explanations and benchmark their performance against recent techniques using proposed evaluation metrics for audio generation tasks.

\textbf{Contributions.} We summarize our contributions as follows: 1) We propose a faithful explanation method for text-to-audio generation models, grounded in factual and counterfactual reasoning to quantify the importance of text tokens to the generated audio. 2) We offer a framework that provides both holistic and granular audio explanations based on user requests, enabling tailored insights. 3) 
We introduce new evaluation metrics for text-to-audio explanations and demonstrate the effectiveness of \mname{}through extensive experiments compared to existing methods. 4) We present case studies demonstrating how \mname{} provides valuable insights to support the understanding of model behavior and editing tasks.
\section{Related Work} \label{sec:related}

\textbf{Text-to-Audio Generation Models.} Recent text-to-audio generation models can be categorized into two model architectures: Transformer-based~\cite{kreuk2022audiogen, ziv2024masked} and Diffusion-based~\cite{yang2023diffsound, liu2023audioldm, schneider2023mo}. Transformer models, such as AudioGen~\cite{kreuk2022audiogen}, employ autoregressive Transformers to predict discrete audio tokens, while MAGNeT~\cite{ziv2024masked} enhances efficiency through masked generative modeling in a non-autoregressive scheme. Diffusion-based approaches such as Diffsound~\cite{yang2023diffsound} generate discrete mel-spectrogram tokens, whereas models like AudioLDM~\cite{liu2023audioldm} and Moûsai~\cite{schneider2023mo} directly predict continuous mel-spectrograms or waveforms. Despite architectural differences, these models commonly use cross-attention mechanisms, making \mname{} a model-agnostic explainer for TAG models that use cross-attention in audio generation.

\textbf{Explainable AI.}  Explainability involves methods that help to understand the importance of each input token with respect to output predictions. These methods generally fall into two categories: gradient-based methods~\cite{selvaraju2017grad, sundararajan2017axiomatic, nagahisarchoghaei2023empirical} and perturbation-based methods~\cite{ribeiro2016should, lundberg2017unified}. Gradient-based explanation methods trace gradients from the target layers to the predictive value, using the calculated gradients as a measure of importance. While effective, these methods require substantial memory resources to store the values of each targeted layer. In contrast, perturbation-based methods, such as SHAP~\cite{lundberg2017unified}, are more memory-efficient, calculating feature importance by comparing predictions with and without specific features. Similarly, our method adopts a perturbation-based approach to effectively generate explanations.

\textbf{Explainability on Audio Processing Models.} Existing explainability approaches~\cite{akman2024audio} on audio processing models have extended generic explanation methods. For instance, one study~\cite{becker2018interpreting} employs Layer-wise Relevance Propagation (LRP) to explain the model trained on raw waveforms and spectrograms for spoken digit and speaker gender classification. Another study applied DFT-LRP~\cite{frommholz2023xai} to audio event detection, assessing the significance of time-frequency components and guiding input representation choices. Similarly, audioLIME~\cite{haunschmid2020audiolime} extends LIME~\cite{ribeiro2016should} to explain music-tagging models by perturbing audio components derived from source separation. However, since the above methods focus on explaining audio continuously and sequentially, they are not directly applicable to the unique challenges posed by TAG models, which require techniques that address the complex interactions between text inputs and generated audio outputs.

\textbf{Explainability on Transformer.} With the widespread use of Transformers, the demand for explainability has grown. Primarily, Rollout~\cite{abnar2020quantifying} primarily aggregates attention weights in all layers to track information flow but struggles to integrate cross-attention weights in multi-modal models with differing domain dimensionalities. Another recent work~\cite{chefer2021generic} leverages Layer-wise Relevance Propagation (LRP)~\cite{samek2017explainable} to calculate class-specific relevance scores based on gradients of attention weights in self- and cross-attention layers. Nevertheless, AtMan~\cite{deiseroth2023atman} raises the issue of excessive memory usage and introduces a scalable explanation method that employs single-token perturbation to observe the change of loss in the response. While intuitive and memory-efficient for large-scale models, this method is limited in its ability to account for the interrelationship of input tokens.
\section{Preliminaries}
AudioGen~\cite{kreuk2022audiogen}, a representative TAG model, consists of three key components: a text encoder~\cite{raffel2020exploring}, an autoregressive Transformer decoder model~\cite{vaswani2017attention}, and an audio decoder~\cite{defossez2022high}. The Transformer decoder serves as the core model responsible for generating the audio sequence, while the text encoder processes the input text and the audio decoder post-processes the generated audio token sequence into audio. Given a text prompt, it is converted into a tokenized representation vector, denoted as $\textbf{U} = [\textbf{u}_{1}, \dots, \textbf{u}_{L}], \textbf{U} \in \mathbb{R}^{L \times d_{u}}$, where $L$ denotes number of textual tokens and $d_{u}$ represents a dimension of the textual token representation vector. The generated audio can be expressed in a discrete form, as EnCodec~\cite{defossez2022high} converts the audio into either discrete tokens or continuous token representations. The tokenized audio sequence is denoted as $\textbf{Z} = [\textbf{z}_{1}, \dots, \textbf{z}_{T}], \textbf{Z} \in \mathbb{N}^{T \times d_{v}}$, where $T$ denotes the length of the audio sequence and $d_{v}$ indicates the number of codebooks $d_{v}$. In detail, the codebook is a structured set of discrete audio tokens used in multi-stream audio generation to produce high-quality audio. For more comprehensive information on multi-streaming audio generation, we refer to the original AudioGen paper~\cite{kreuk2022audiogen}. 

For the generation of an audio sequence, the Transformer-decoder model ~\cite{vaswani2017attention}, denoted as $h$, generates $\textbf{z}_{t}$ as $t$-th order audio token in the sequence, following the formulation $h(\textbf{U}, \textbf{z}_{t-1}) = \textbf{z}_{t}$. For brevity, we omit the detailed notation of other components and the top-$p$ or top-$k$ sampling process in the Transformer. Instead, we focus on the attention layers, including cross-attention, which are crucial components of the model, denoted as $f$. The computation within these layers is expressed in a simplified version as $f(\textbf{U}, \textbf{z}_{t-1}) = \textbf{e}_{t}$, where $\textbf{e}_{t}$ represents the latent representation vector corresponding to the $t$-th audio token. In the absence of ground truth and class labels, the latent embedding vector $\textbf{e}_{t}$ in the audio token space provides information on how perturbation impacts subsequent generations. Particularly, the cross-attention layer is essential to fuse the textual information with auditory information in layers $f$, we denote the cross-attention layers as:
\begin{equation}
\displaystyle g(\textbf{Q}, \textbf{K}, \textbf{V}) =  \sigma \left(\frac{\textbf{Q} \textbf{K}^\intercal}{\sqrt{d_k}}\right)\textbf{V},
\label{att}
\end{equation}
where $\sigma$ indicates a softmax function, $\textbf{Q}, \textbf{K}, \textbf{V}, d_k$ refers to query, key, values, and the number of vector dimensions in the $k$-th layer, respectively. In detail, $\textbf{Q}$ refers to previously generated audio tokens, representing the query information, while$\textbf{K}$ and $\textbf{V}$ correspond to the textual tokens.

\section{The Proposed \mname{}}
\begin{figure*}[!ht]
    \center
    \includegraphics[width=1.0 \linewidth]{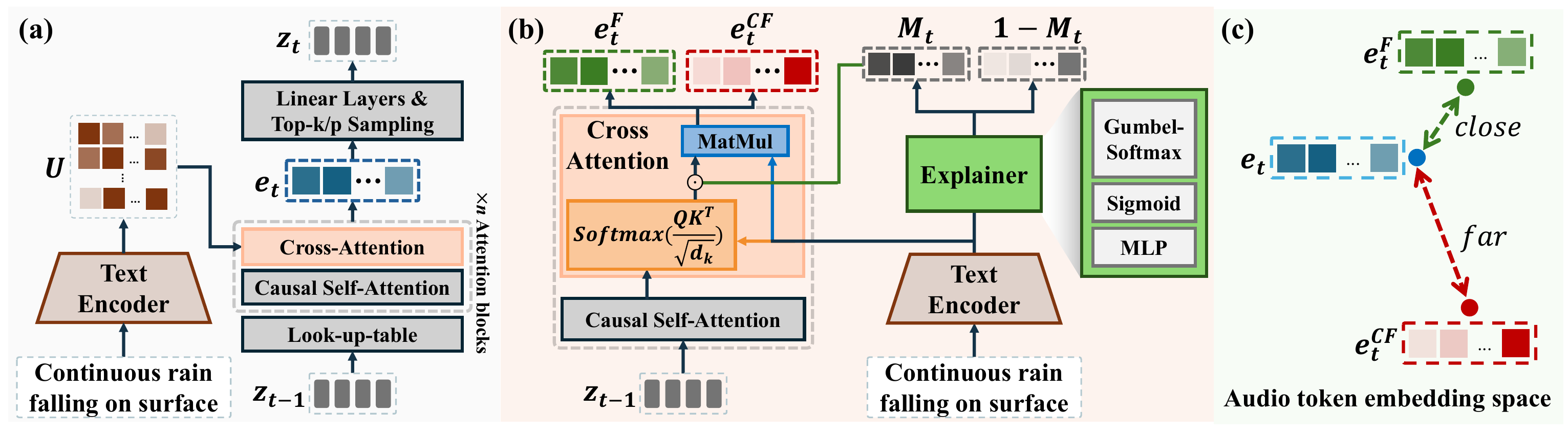}
    \caption{(a) The process by which AudioGen generates an audio. (b) \mname{}'s procedure for generating and applying explanations, with the $Explainer$ in the green box. (c) The method for calculating and applying the loss in \mname{}.}
    \label{overview}
\end{figure*}

\mname{} addresses the challenge of explaining TAG models, where the goal is to quantify the importance of textual input corresponding to the generated audio. To achieve this within a sequence-to-sequence framework, we decompose the explanation target, represented as sequential audio, into individually non-sequential audio tokens. Since the output is sequential data, calculating gradients across the entire sequence, from the first to the last token, is computationally expensive and time-consuming. To overcome these issues, we redefine the explanation target as individual audio tokens, rather than the entire sequence. This modification enables parallel computation of generating an explanation for each token, significantly speeding up the process. Finally, \mname{} integrates these individual token-level explanations to provide a comprehensive understanding of the entire audio sequence. An overview of \mname{} is illustrated in Figure~\ref{overview}.

\subsection{Definition of Masks as Explanations} 
We quantify the importance of the $t$-th audio token $\textbf{z}_{t}$ within the audio sequence using a mask as the explanation. The soft mask is denoted as $\textbf{M}_{\textbf{U}, \textbf{z}_{t}} \in \mathbb{R}^{L \times 1}$, where each element $\textbf{m}_{\textbf{u}_{i}, \textbf{z}_{t}} \in \textbf{M}_{\textbf{U}, \textbf{z}_{t}}$ represents the importance of the $i$-th textual token with respect to the $t$-th audio token $\textbf{z}_{t}$. Each value lies in the range $[0, 1]$, where a value close to 1 indicates that the corresponding textual token is highly important for generating the target audio token, while a value closer to 0 indicates lower importance. To serve as a soft mask representing the importance of each text token, \mname{} optimizes the $Explainer$ to predict the mask $\textbf{M}_{\textbf{U}, \textbf{z}_{t}}$ as the explanation. The $Explainer$ consists of Multi-Layer Perceptrons (MLPs) with a sigmoid and gumbel-softmax~\cite{jang2016categorical} function to constrain values within the range $[0, 1]$ without additional scaling and to enforce the values close to either $0$ or $1$, thereby highlighting relatively distinguished contribution. Using the soft mask, we apply perturbation to modify the inner computational steps of the cross-attention layers, altering the attention score of the given textual input. Consequently, we measure the perturbation effect on the prediction at the layer $f(\textbf{U}, \textbf{z}_{t-1}) = \textbf{e}_{t}$, observing how latent representation vector $\textbf{e}_{t}$ for the audio token $\textbf{z}_{t}$ changes under these perturbations. In the following section, we detail how we optimize $Explainer$ to predict the mask as explanations based on both factual and counterfactual reasoning.

\subsection{Formulating Factual Explanations}

Factual reasoning ~\cite{tan2022learning, ali2023explainable,kenny2021post} aims to find sufficient input that can approximately reproduce the original prediction. To quantify the sufficiency of textual tokens, we employ a perturbation-based method using the soft mask, interleaving the computation to measure the impact of changes. Specifically, we mask out attention scores in the cross-attention layers where textual information is fed into the TAG model. We formulate the perturbation in factual reasoning as:
\begin{equation}
\displaystyle \tilde{g}(\textbf{Q}, \textbf{K}, \textbf{V}, \textbf{M}_{\textbf{U}, \textbf{z}_{t}}) =  (\sigma(\frac{\textbf{Q}\textbf{K}^\intercal}{\sqrt{d_k}}) \odot \textbf{M}_{\textbf{U}, \textbf{z}_{t}}) \textbf{V},
\label{attx}
\end{equation}
\noindent where $\sigma$ denotes the softmax activation function and the mask $\textbf{M}_{\textbf{U}, \textbf{z}_{t}}$ controls the amount of information corresponding to the text token. When the mask value $\textbf{m}_{\textbf{u}_{i}, \textbf{z}_{t}}$ approaches $0$, the attention score is suppressed, meaning the information corresponding to the textual token is not fully propagated to the subsequent layer. Conversely, as the mask value approaches $1$, the original value is fully preserved. To distinguish this process from the original layer $f(\textbf{U}, \textbf{z}_{t-1})$, we denote the layer applying perturbation with the factual mask as $\tilde{f}(\textbf{U}, \textbf{z}_{t-1}, \textbf{M}_{\textbf{U}, \textbf{z}_{t}})$.

When the mask sufficiently serves as a factual explanation, the perturbed output remains approximately the same as the original prediction. To evaluate the impact of perturbation, we measure the resulting changes in the latent representation vector within the audio token space. Since the latent vector encodes rich and implicit information, we expect that two vectors close to each other indicate a similar auditory meaning, which is likely to result in similar audio generation. By leveraging this vector similarity, we can effectively measure the influence of perturbation and formulate the objective function for $Explainer$ as:

\begin{equation}
\mathcal{L}_{F} = -cos(f(\textbf{U}, \textbf{z}_{t-1}), \tilde{f}(\textbf{U}, \textbf{z}_{t-1}, \textbf{M}_{\textbf{U}, \textbf{z}_{t}})),
\label{fact}
\end{equation}
\noindent where $cos$ function refers to cosine similarity, which measures how similar factual result $\tilde{f}(\textbf{U}, \textbf{z}_{t-1}, \textbf{M}_{\textbf{U}, \textbf{z}_{t}})$ is to the original prediction $f(\textbf{U}, \textbf{z}_{t-1})$ in the audio token space. Since the objective function involves negative cosine similarity, minimizing the loss function corresponds to maximizing the similarity. Hence, following the objective function, the $Explainer$ generates the factual explanation mask, ensuring that the two representations or predictions are as close as possible in the audio token space.

\subsection{Formulating Counterfactual Explanations}
Counterfactual reasoning~\cite{tan2022learning, ali2023explainable, kenny2021post} aims to identify necessary inputs that can significantly alter the original prediction when it is perturbed or removed. This perturbation operates in the opposite direction of factual explanations, removing the important input to observe the counterfactual result. We formulate the perturbation method in counterfactual reasoning as:
\begin{equation}
\displaystyle \tilde{g}(\textbf{Q}, \textbf{K}, \textbf{V}, \textbf{1}-\textbf{M}_{\textbf{U}, \textbf{z}_{t}}) =  (\sigma(\frac{\textbf{Q}\textbf{K}^\intercal}{\sqrt{d_k}}) \odot (\textbf{1}-\textbf{M}_{\textbf{U}, \textbf{z}_{t}})) \textbf{V},
\label{att_cf}
\end{equation}
\noindent where $\textbf{1} \in \mathbb{R}^{1 \times T_{u}}$ is a vector of ones and $\textbf{1}-\textbf{M}_{\textbf{U}, \textbf{z}_{t}}$ subtracts the importance of the corresponding textual tokens. Consequently, the more important a textual token is, the more its attention score is suppressed in proportion to its importance. This perturbation operates under a counterfactual assumption as the What-If scenario~\cite{tan2022learning, ali2023explainable, kenny2021post}: What happens if the important textual token does not exist? After applying the perturbation in Equation~\eqref{att_cf}, the counterfactual result is observed as $\tilde{f}(\textbf{U}, \textbf{z}_{t-1}, \textbf{1}-\textbf{M}_{\textbf{U}, \textbf{z}_{t}})$. If the counterfactual result significantly differs from the original prediction, it indicates that the counterfactual mask is necessary to explain the original prediction. Conversely, if the change is trivial, the counterfactual mask is unnecessary to explain the causal relationship with the prediction. 

Generally, counterfactual explanations in supervised settings aim to find the important inputs that change the prediction with minimal perturbation. However, no class labels or guidance are available in our task of audio generation. Instead, we measure the change of meaning in latent space leveraging the cosine similarity function after counterfactual perturbation. Thus, the counterfactual explanation objective function is formulated as:
\begin{equation}
    \mathcal{L}_{CF} = cos(f(\textbf{U}, \textbf{z}_{t-1}), \tilde{f}(\textbf{U}, \textbf{z}_{t-1}, \textbf{1}-\textbf{M}_{\textbf{U}, \textbf{z}_{t}})),
    \label{cfact}
\end{equation}
\noindent where $cos$ function measures how dissimilar counterfactual result $\tilde{f}(\textbf{U}, \textbf{z}_{t-1}, \textbf{1}-\textbf{M}_{\textbf{U}, \textbf{z}_{t}})$ is to the original prediction in latent space. As the cosine similarity decreases, the objective function minimizes the similarity. Consequently, the $Explainer$ generates the counterfactual explanation mask to ensure that the two representations or predictions are as far as possible in the audio token space after counterfactual perturbation.

\subsection{Objective Function for \mname{}} 
Along with factual and counterfactual explanation objective functions, we add the regularization term to generate the explanation mask in a simple and efficient manner. Therefore, we incorporate additional regularization in our final objective function for the $Explainer$, which is formulated as:
\begin{equation}
    \mathcal{L} = \mathcal{L}_{F} + \mathcal{L}_{CF} + \alpha \cdot L_1(\textbf{M}_{\textbf{U}, \textbf{z}_{t}})+ \beta \cdot L_2(\textbf{M}_{\textbf{U}, \textbf{z}_{t}}).
    \label{total}
\end{equation}
Here, $L_1$ and $L_2$ represent the $L_1$-Norm and $L_2$-Norm, respectively, as regularization terms to minimize the mask size. This prevents a trivial solution where the $Explainer$ generates an explanation mask assigning equal importance to all values. At the same time, adhering to Occam’s Razor principle, we favor simpler and more effective explanations~\cite{tan2022learning, blumer1987occam}. Hence, according to the objective function in Equation~\eqref{total}, \mname{} optimizes the $Explainer$ generating faithful explanation masks in the audio-token level.

\begin{algorithm}[!t]
\caption{\mname{}}
\begin{algorithmic}
    \State \textbf{Input}: Textual token representation vector \textbf{U}, \\
    previously generated audio token vector $\textbf{z}_{t-1}$, Transformer model $f$, audio generation length $T$, number of epochs $K$, learning rate $\lambda$, regularization coefficients $\alpha$ and $\beta$ 
    \For {$t = 1$ \textbf{to} $T$}
        \State \textbf{Initialize} $Explainer$ with random parameters.
        
            \For {$epoch = 1$ \textbf{to} $K$}
                \State $\textbf{M}_{\textbf{U}, \textbf{z}_{t}} = Explainer(\textbf{U}, \textbf{z}_{t-1})$
                \State $L = L_{F} +  L_{CF} + \alpha \cdot L_1(\textbf{M}_{\textbf{U}, \textbf{z}_{t}}) + \beta \cdot L_2(\textbf{M}_{\textbf{U}, \textbf{z}_{t}})$
                \State $\theta := \theta - \lambda \nabla_\theta L$
            \EndFor
        \State $\textbf{M}_{\textbf{U}, \textbf{z}_{t}} = Explainer(\textbf{U}, \textbf{z}_{t-1})$
    \EndFor
    \State \textbf{Return} $\textbf{M}_{\textbf{U}, \textbf{z}} =  \displaystyle \frac{1}{T} \sum_{t=1}^{T} \textbf{M}_{\textbf{U}, \textbf{z}_{t}}$
\end{algorithmic}
\end{algorithm}

\subsection{Providing Audio-Level Explanations}
In this section, we aggregate audio token-level explanations to provide a comprehensive understanding of the entire audio sequence. The aggregation is performed by averaging the mask values across all audio tokens as follows:
\begin{equation}
\textbf{M}_{\textbf{U}, \textbf{z}} =  \displaystyle \frac{1}{T} \sum_{t=1}^{T} \textbf{M}_{\textbf{U}, \textbf{z}_{t}},
\label{audioexpl}
\end{equation}
\noindent where $t$ refers to the step, and $T$ represents the total length of generated audio. Additionally, it is possible to focus on a specific interval of interest within the audio, defined between a starting step $s$ and an ending step $n$. This is denoted as $\textbf{M}_{\textbf{U}, \textbf{z}} = \frac{1}{|n-s|+1} \sum_{t=s}^{n} \textbf{M}_{\textbf{U}, \textbf{z}_{t}}$, which provides granular explanations based on the user's request. This flexible approach enables users to discover patterns within specific intervals, as \mname{} can effectively capture and explain auditory content in targeted regions of the audio sequence.
\section{Experimental Setup}
\label{exp-setup}

\textbf{Dataset.} We use AudioCaps~\cite{kim2019audiocaps} as the source of textual prompts. For each prompt, we generate a 5-second audio clip using AudioGen, pairing each prompt with its corresponding generated audio. For hyperparameter tuning, we select 100 validation captions, while the test dataset consists of 1,000 randomly selected captions.

\textbf{Evaluation Metrics.} We evaluate explanations based on two metrics: Fidelity and KL divergence, both derived from the classification probabilities of a pre-trained audio classifier. Specifically, we utilize PaSST~\cite{cai2022efficient}, a classifier trained on the AudioSet dataset, which is also used in the evaluation of AudioGen. Its classification probabilities are likely to provide meaningful insights into the relationship between textual prompts and generated audio. Fidelity~\cite{yuan2021explainability, ali2023explainable}, a core evaluation metric in the field of XAI, measures the change in top-1 label prediction probabilities of the generated audio after applying factual and counterfactual explanation masks, denoted as $Fid_{F}$ and $Fid_{CF}$, respectively. 

In addition, KL divergence~\cite{kilgour2018fr}, originally used to evaluate audio generative models~\cite{kreuk2022audiogen, yang2023diffsound, huang2023make}, measures the differences of label distribution between generated and reference audio. For explanation evaluation, we introduce new metrics $KL_{F}$ and $KL_{CF}$, which measure the conceptual change in the generated audio after applying explanation masks in factual and counterfactual reasoning, respectively. In factual evaluation, the generated audio should closely match the original audio, making lower values $Fid_{F}$ and $KL_{F}$ desirable. In contrast, in counterfactual evaluation, higher values of $Fid_{CF}$ and $KL_{CF}$ indicate a more effective explanation. Additionally, we include the average mask size as part of our evaluation.

\textbf{Baselines.} We compare our method with five baselines. Random-Mask is a mask with randomly assigned values ranging between 0 and 1. Grad-CAM~\cite{selvaraju2017grad} is evaluated in two variations: $\text{Grad-CAM-a}$ and $\text{Grad-CAM-e}$. Specifically, $\text{Grad-CAM-a}$ computes the gradients of the latent representation vector of the $t$-th audio token $\textbf{e}_t$ with respect to the generated audio sequence $\textbf{z}_t$, while $\text{Grad-CAM-e}$ computes the gradients of the last cross-attention map to the $\textbf{z}_t$. We also include the AtMan~\cite{deiseroth2023atman} and the method proposed by Chefer et al.~\cite{chefer2021generic} as baselines.

\textbf{Experimental Setting.} The $Explainer$ model includes a linear layer that reduces the text token embeddings from 1536 to 512 dimensions, followed by a PReLU activation function. The 512-dimensional text token embeddings are then mapped to a single value through another linear layer and a sigmoid function, producing a value in the $[0, 1]$ range. A Gumbel-Softmax function is subsequently applied to push values closer to 0 or 1, representing the importance of each text token. The $Explainer$ is trained for 50 epochs with a learning rate as $\times 10^{-3}$ using the Adam optimizer. Hyperparameters are set as $\alpha = 1 \times 10^{-3}$ and $\beta = 1 \times 10^{-1}$ as coefficients for the explanation objective function. Hyperparameter sensitivity analysis and detailed experimental settings are both provided in the Appendix. Our code is available at the following link \footnote{\url{https://github.com/hjkng/audiogenX}}.

\begin{table*}[!ht]
\centering
\begin{tabular}{llllll}
\toprule
Method& $Fid_{F}\downarrow$ & $Fid_{CF}\uparrow$ & $KL_{F}\downarrow$ & $KL_{CF}\uparrow$ & Size $\downarrow$ \\
\midrule
    $N_{audio}=5$ & $0.128 \pm 0.004$ & \multicolumn{1}{c}{-} & $1.318 \pm 0.030$ & \multicolumn{1}{c}{-} & \multicolumn{1}{c}{-}\\
    Random-Mask & $0.196 \pm 0.004$ & $0.195 \pm 0.006$ & $1.884 \pm 0.044$ & $1.932 \pm 0.046$ & 0.500 \\
    $\text{Grad-CAM-e}$ & $0.204 \pm 0.006$ & $0.235 \pm 0.008$ & $1.858 \pm 0.034$ & $2.457 \pm 0.041$ & 0.422 \\
    $\text{Grad-CAM-a}$ & $0.240 \pm 0.006$ & $0.192 \pm 0.010$ & $2.285 \pm 0.077$ & $1.951 \pm 0.075$ & 0.406 \\
    AtMan & $0.195 \pm 0.008$ & $0.222 \pm 0.008$ & $2.010 \pm 0.049$ & $2.198 \pm 0.048$ & 0.497 \\
    Chefer et al. & $0.198 \pm 0.003$ & $0.229 \pm 0.004$ & $1.899 \pm 0.025$ & $2.348 \pm 0.040$ & 0.441 \\ \hline
    \mname \ w/ Eq.~\eqref{fact} & $0.145 \pm 0.004$ & $0.360 \pm 0.005$ & $1.542 \pm 0.024$ & $3.658 \pm 0.061$ & \textbf{0.360} \\
    \mname \ w/ Eq.~\eqref{cfact} & $0.143 \pm 0.004$ & $0.385 \pm 0.005$ & $1.514 \pm 0.043$ & $3.977 \pm 0.044$ & 0.385 \\
    \mname \ w/ Eq.~\eqref{audioexpl} & $0.137 \pm 0.005$ & $0.402 \pm 0.005$ & $1.418 \pm 0.043$ & $4.183 \pm 0.073$ & 0.455 \\
    \mname & $\textbf{0.132} \pm 0.004$ & $\textbf{0.405} \pm 0.004$ & $\textbf{1.416} \pm 0.029$ & $\textbf{4.259} \pm 0.039$ & 0.455 \\
\bottomrule
\end{tabular}
\caption{Evaluation of explanations generated by each method using factual and counterfactual reasoning. Five audio samples are generated and evaluated with different seeds based on the obtained explanations. The best results are highlighted in \textbf{bold}.}
\label{table:main}
\end{table*}
\section{Experimental Results}

\begin{figure}
    \center
    \includegraphics[width=1.0 \linewidth]{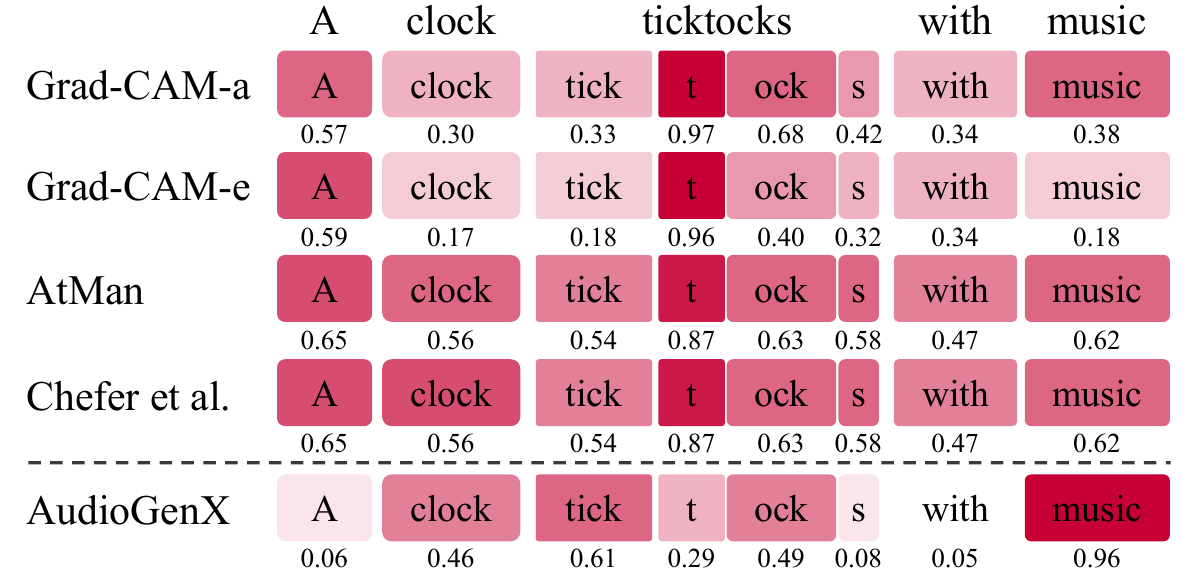}
    \caption{Visualization of \mname{} and other methods.}
    \label{experiment}
\end{figure}

\subsection{RQ 1: Does \mname{} Generate Faithful Explanations?}

We evaluate the generated explanations by \mname{} based on factual and counterfactual reasoning, as presented in Table~\ref{table:main}. \mname{} achieves the best performance across the metrics $Fid_{F},$ $Fid_{CF}$, $KL_{F}$, and $KL_{CF}$, while also maintaining the smallest size ($Size$), demonstrating that our explanations are both simple and effective. The baseline, denoted as $N_{audio}=5$, generates audio conditioned on the same textual input five times to observe the inherent variance, serving as the lower bound for $Fid_{F}$, $KL_{F}$. \mname{}'s factual audio nearly reaches the lower bound, indicating high performance. Furthermore, significant changes in $Fid_{CF}$ and $KL_{CF}$ under counterfactual perturbations confirm that the explanations are both sufficient and necessary. The \mname{} with factual and counterfactual losses in Eq.\eqref{total}, outperforms the variants \mname \ w/ Eq.~\eqref{fact} and \mname \ w/ Eq.~\eqref{cfact}, which apply only factual or counterfactual loss with a regularization term. This indicates that the two losses complement each other, enhancing overall performance. Furthermore, we evaluate \mname \ w/ Eq.~\eqref{audioexpl} using an averaged explanation mask, showing the robustness of explainability in describing the entire audio. In contrast, other baselines fail to generate meaningful counterfactual audio, lacking the optimization properties needed to enforce counterfactual explanations. 

The strong performance highlights the effectiveness of leveraging latent embedding vectors to generate explanations. While most baselines are designed to explain supervised learning models, they rely on vectors that represent the probability distribution of the final audio token. This approach, however, does not align well with the inference process of audio generation models. In extreme cases, such as top-$k$ sampling ($k$=250), the 250-th audio token could be sampled, leading to significant discrepancies between the gradients or probability-related information the token most likely predicted by the model. In contrast, our approach avoids dependency on the sampling process, allowing the model to produce more faithful explanations.

\subsection{RQ 2: How Well Do the Explanations from \mname{} Reflect the Generated Audio?}

We visualize the explanations generated by \mname{} and other baselines, as shown in Figure~\ref{experiment}. \mname{} demonstrates a clear advantage in focusing on key audio elements. Unlike other baselines, which often assign relatively high importance scores to less important tokens like `A' and `with', \mname{} consistently assigns higher importance scores to crucial tokens such as `ticktocks' and `music'. For instance, \mname{} assigns a notably high importance score of 0.96 to `music', emphasizing its ability to focus on significant input tokens. In contrast, other models like $\text{Grad-CAM-e}$ and AtMan distribute importance more broadly, including less relevant tokens. These results show that \mname{} consistently provides faithful explanations, aligning the generated audio with the essential components of the input text.

\begin{figure}
    \center
    \includegraphics[width=1.0 \linewidth]{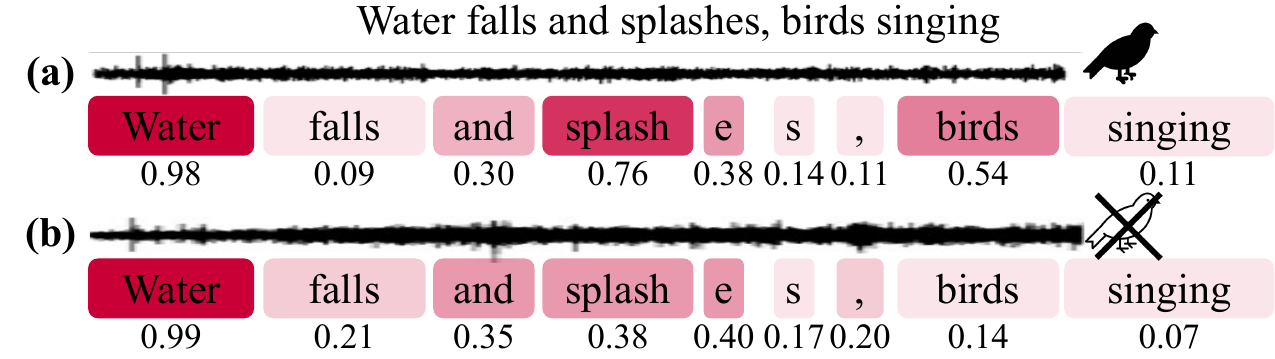}
    \caption{Explanation generated by \mname{} for two audios created from a single prompt. (a) includes bird sounds, while (b) does not.}
    \label{rq2}
\end{figure}

Furthermore, when generating audio from a prompt containing multiple concepts, some words may be less prominently reflected. In such case, \mname{} provides adequate explanations for each specific audio, indicating whether each word from the prompt has been incorporated into the generated audio. As illustrated in Figure~\ref{rq2}, the difference between the two audios is that bird sounds are present in Figure~\ref{rq2}-(a) but absent in Figure~\ref{rq2}-(b).
\mname{} effectively describes the audios by assigning high importance scores of 0.98 and 0.99 to the token `Water,' which is the primary sound in both audios. \mname{} assigns a score of 0.54 to `birds,' while it assigns a score of 0.14, accurately reflecting the different audio characteristics in each case. These results show that \mname{} can provide explanations that are well-suited to the corresponding audio. Furthermore, these explanations serve as valuable insights for editing generated audio to better align with user intention. 

\subsection{RQ 3: How Can Explanations Help Understand AudioGen Behavior?} 

\begin{figure}
    \center
    \includegraphics[width=1.0 \linewidth]{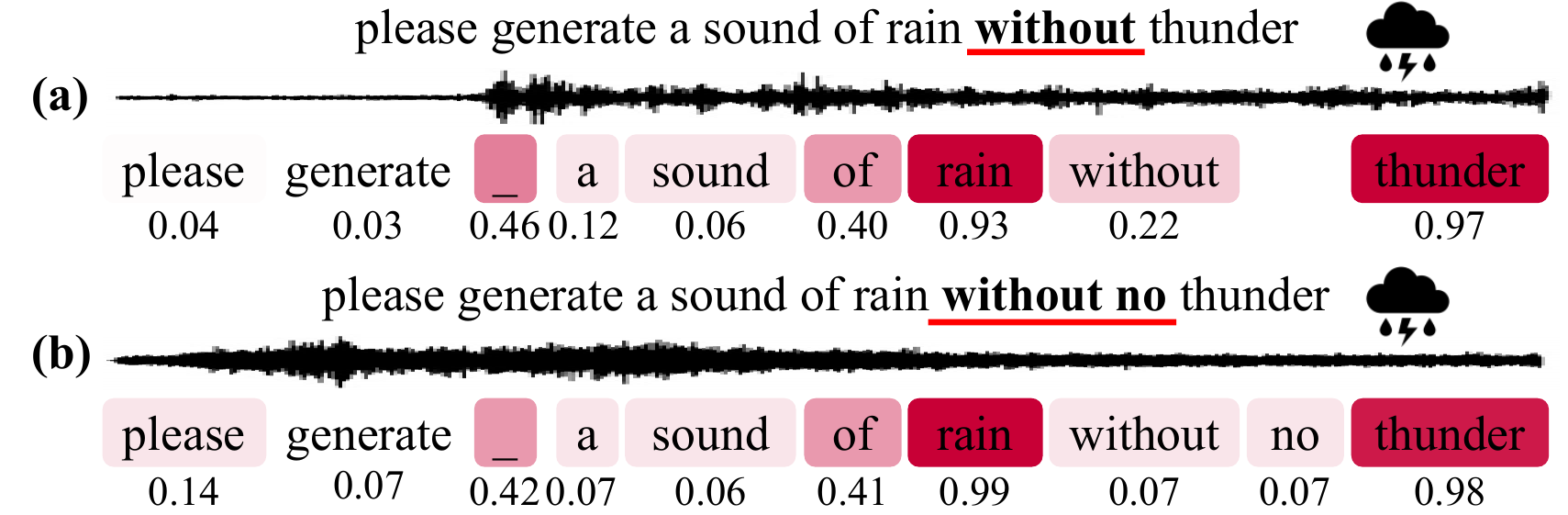}
    \caption{Explanations generated from negated prompts: (a) single negation, (b) double negation.}
    \label{rq3}
\end{figure}

We explore the output patterns of AudioGen using the explanations generated by \mname{}. First, we investigate whether AudioGen can effectively handle sentences containing negations and double negations, as shown in Figure~\ref{rq3}. The explanations of the generated audios are presented in response to input prompts containing `without thunder' and `without no thunder.' In both cases, the generated audio includes the sound of thunder along with the rain. Using \mname{}, we observe that `without' and `without no' have lower importance compared to 'thunder' in the explanations. We hypothesize that this occurs because the training dataset lacks sufficient examples of negation and double negation. An examination of the AudioCaps dataset reveals a scarcity of such cases. Additionally, by aggregating tokens from the explanations, we identify the top and bottom 50 tokens in Table~\ref{tabler3} in the Appendix. Tokens with high importance are predominantly nouns, such as `thunder,' while those with low importance include sound descriptors like `distant,' as well as sequential expressions like `before.' Such analyses could be used to debug TAG models or to identify potential inherent biases in their behavior. 

\subsection{RQ 4: Does \mname{} Generate Explanations Efficiently?}

\begin{table}[!ht]
\centering
\begin{tabular}{lrr}
\toprule
Method& Memory (MB)$\downarrow$& Time (s)$\downarrow$ \\
\midrule
    $\text{Grad-CAM-e}$ & 8641.306 & 49.038  \\
    $\text{Grad-CAM-a}$ & 41655.848 & 62.276  \\
    AtMan & \textbf{5081.957} & \textbf{7.295}  \\
    Chefer et al. & 41684.969 & 52.166 \\ \hline
    \mname \ w/ Eq.~\eqref{fact} & 11980.894 & 36.639 \\
    \mname \ w/ Eq.~\eqref{cfact} & 11981.114 & 37.373 \\
    \mname \ w/ Eq.~\eqref{audioexpl} & 12001.931 & 63.198 \\
    \mname & 12001.931 & 63.198 \\
\bottomrule
\end{tabular}
\caption{Efficiency analysis of \mname{} and other baseline methods. The best results are highlighted in \textbf{bold}.}
\label{table:eff}
\end{table}

We evaluate the efficiency of explanation methods based on the average time and total GPU memory usage per explanation, as shown in Table~\ref{table:eff}. For GPU memory efficiency, the results rank in the following order: AtMan, $\text{Grad-CAM-e}$, \mname{}, $\text{Grad-CAM-a}$, and Chefer et al. For time efficiency, the order is AtMan, $\text{Grad-CAM-e}$, Chefer et al., $\text{Grad-CAM-a}$, and \mname{}. Although AtMan is the most efficient, its performance remains subpar due to its simplistic approach. $\text{Grad-CAM-e}$ demonstrates greater memory efficiency compared to $\text{Grad-CAM-a}$ and Chefer et al., as it tracks a shallower layer. While \mname{} requires additional computational time to train explanation masks, it achieves memory efficiency by reducing GPU storage and operates with $\mathcal{O}(Lk)$ complexity, ensuring linear scalability for large-scale tasks.
\section{Conclusion}
\mname{} quantifies the importance of textual tokens corresponding to generated audio by leveraging both factual and counterfactual reasoning frameworks. This approach enables the generation of faithful explanations, providing actionable insights for users to edit audio and assisting developers in debugging. Consequently, \mname{} enhances the transparency and trustworthiness of TAG models.

\section{Acknowledgements}
This work was supported by NCSOFT, the Institute of Information \& Communications Technology Planning \& evaluation (IITP) grant, and the National Research Foundation of Korea (NRF) grant funded by the Korean government (MSIT) (RS-2019-II190421, IITP-2025-RS-2020-II201821, RS-2024-00438686, RS-2024-00436936, RS-2024-00360227, RS-2023-0022544, NRF-2021M3H4A1A02056037, RS-2024-00448809).  This research was also partially supported by the Culture, Sports, and Tourism R\&D Program through the Korea Creative Content Agency grant funded by the Ministry of Culture, Sports and Tourism in 2024 (RS-2024-00333068, RS-2024-00348469 (25\%)).
\newpage

\bibliography{reference}
\clearpage
\section{Appendix}

\subsection{Evaluation metrics}
We detail the four evaluation metrics discussed in the main manuscript. $\textbf{Z}_i$ represents the audio token sequence generated by a text-to-audio model given the $i$-th text prompt in the dataset. The sequences $(\tilde{\textbf{Z}}_{F})_{i}$ and $(\tilde{\textbf{Z}}_{CF})_{i}$ are produced by applying factual and counterfactual masks, respectively, as defined in Eqs.~\eqref{attx} and \eqref{att_cf}. These masks serve as explanations for both $\textbf{Z}_i$ and its associated text prompt.
To evaluate these explanations, we use the pre-trained audio classifier PaSST~\cite{koutini2021efficient}, denoted as $q$, which generates a prediction probability distribution $q(\textbf{Z})$. Specifically, $q(\textbf{Z})_{y_i}$ represents the prediction probability for class $y_i$. Finally, $N$ is the total number of data points, $L$ is the number of text tokens in the text prompt, and $T$ is the total length of the audio token sequence. The evaluation metrics $Fid_{F}$, $Fid_{CF}$~\cite{yuan2021explainability, yuan2022explainability, ali2023explainable} are defined as:
\begin{equation}
Fid_{F} =  \frac{1}{N}\sum^{N}_{i=1}q(\textbf{Z}_{i})_{y_i} - q((\tilde{\textbf{Z}}_{F})_i)_{y_i},
\end{equation}
\begin{equation}
Fid_{CF} =  \frac{1}{N}\sum^{N}_{i=1}q(\textbf{Z}_i)_{y_i} - q((\tilde{\textbf{Z}}_{CF})_i)_{y_i},
\end{equation}

\noindent where $y_i$ is the predicted class for $q(\textbf{Z}_{i})$, defined as $y_i = \arg\max_{c \in C} q(\textbf{Z}_{i})_c$, and $C$ is the set of all classes that $q$ predicts. Moreover, $KL_{F}$, $KL_{CF}$~\cite{kreuk2022audiogen, yang2023diffsound, huang2023make}, and $Size$ are defined as:
\begin{equation}
KL_{F} = \frac{1}{N}\sum^{N}_{i=1} D_{\text{KL}} \left( q(\textbf{Z}_{i}) \, \| \, q((\tilde{\textbf{Z}}_{F})_i) \right),
\end{equation}
\begin{equation}
KL_{CF} = \frac{1}{N}\sum^{N}_{i=1}D_{\text{KL}} \left( q(\textbf{Z}_i) \, \| \, q((\tilde{\textbf{Z}}_{CF})_i) \right),
\end{equation}
\begin{equation}
    Size = \frac{1}{N \times L \times T} \sum_{i=1}^{N} \sum_{l=1}^{L} \sum_{t=1}^{T} (\textbf{m}_{\textbf{u}_{l}, \textbf{z}_{t}})_{i}.
\end{equation}
Here, $D_{KL}$ refers to the KL divergence.
bsection{Details on baseline setting}
While there is no existing XAI model specifically designed for explaining text-to-audio generative models, we adopt Transformer explanation methods for evaluation.

\textbf{Rollout}~\cite{abnar2020quantifying}, a method for explaining Transformers, proposes aggregating attention scores recursively by multiplying attention maps in all layers. The proposed method named rolling out the attention weights is formulated as below:
\begin{equation}
\tilde{A}(l_i) = 
\begin{cases} 
A(l_i)\tilde{A}(l_{i-1}) & \text{if } i > j \\ 
A(l_i) & \text{if } i = j,
\end{cases}
\end{equation}
\noindent where $\tilde{A}$ means attention rollout. $A(l_i)$ represents ${i}$-th raw attention map. However, applying Rollout in models with cross-attention blocks designed to handle multi-modality is challenging because the dimensions of the attention maps do not match. Therefore, we exclude Rollout from our baselines. 

\textbf{Grad-CAM}~\cite{selvaraju2017grad} computes the gradients of the output activations from the target layer with respect to the final prediction. The importance map is then calculated as:

\begin{equation}
\tilde{A} = \mathbb{E}((\nabla A \odot A)^{+}),
\end{equation}
where $A$ represents the output activations of the target layer, and $\nabla A$ represents the gradient of these activations with respect to the prediction. Specifically, $\nabla A$ comprises the gradients computed for each selected codebook, which are averaged afterward. $\odot$ denotes element-wise multiplication, and $(\cdot)^{+}$ extracts positive values. 
$\textbf{Grad-CAM-a}$ calculates gradient of the latent representation vector $\textbf{e}_t$, corresponding to the $t$-th audio token. The 1,536 dimensions of the audio token embedding are treated as separate channels, and their mean is used to compute the token importance score. $\textbf{Grad-CAM-e}$ derives the gradient of the last cross-attention map, and their mean is used as the token importance score.

\textbf{AtMan}~\cite{deiseroth2023atman} extracts important tokens by perturbation of a single token. For all cross-attention layers and all heads, we multiply \(1-k\) with the pre-softmax attention scores for the target text tokens to introduce perturbation. The value $k$ is consistently set to 0.9, following the configuration in~\cite{deiseroth2023atman}. To quantify the influence of tokens, we calculate the difference in cross-entropy for each codebook and use the sum of these differences as the token importance.

\textbf{Chefer et al.}~\cite{chefer2021generic} calculates the relevance score, following attention layers. In our experiments, we follow ~\cite{chefer2021generic}. However, since we do not consider the influence of the text encoder, we replace it with an identity matrix.

To scale importance values between 0 and 1, we apply Max scaling for each sequence for all baselines except AtMan, For AtMan, which includes negative values, we use Min-Max scaling.

\subsection{Experimental Setting}
In our experiments, we used the following packages and hardware:

\begin{itemize}
    \item \texttt{Python 3.9.18}
    \item \texttt{spacy==3.5.2}
    \item \texttt{torch>=2.1.0}
    \item \texttt{torchaudio>=2.1.0}
    \item \texttt{Transformers>=4.31.0}
\end{itemize}

All computations were performed using a single NVIDIA A100 GPU.

\subsection{Hyperparameter Sensitivity Analysis}
\begin{figure}
    \center
    \includegraphics[width=1.0 \linewidth]{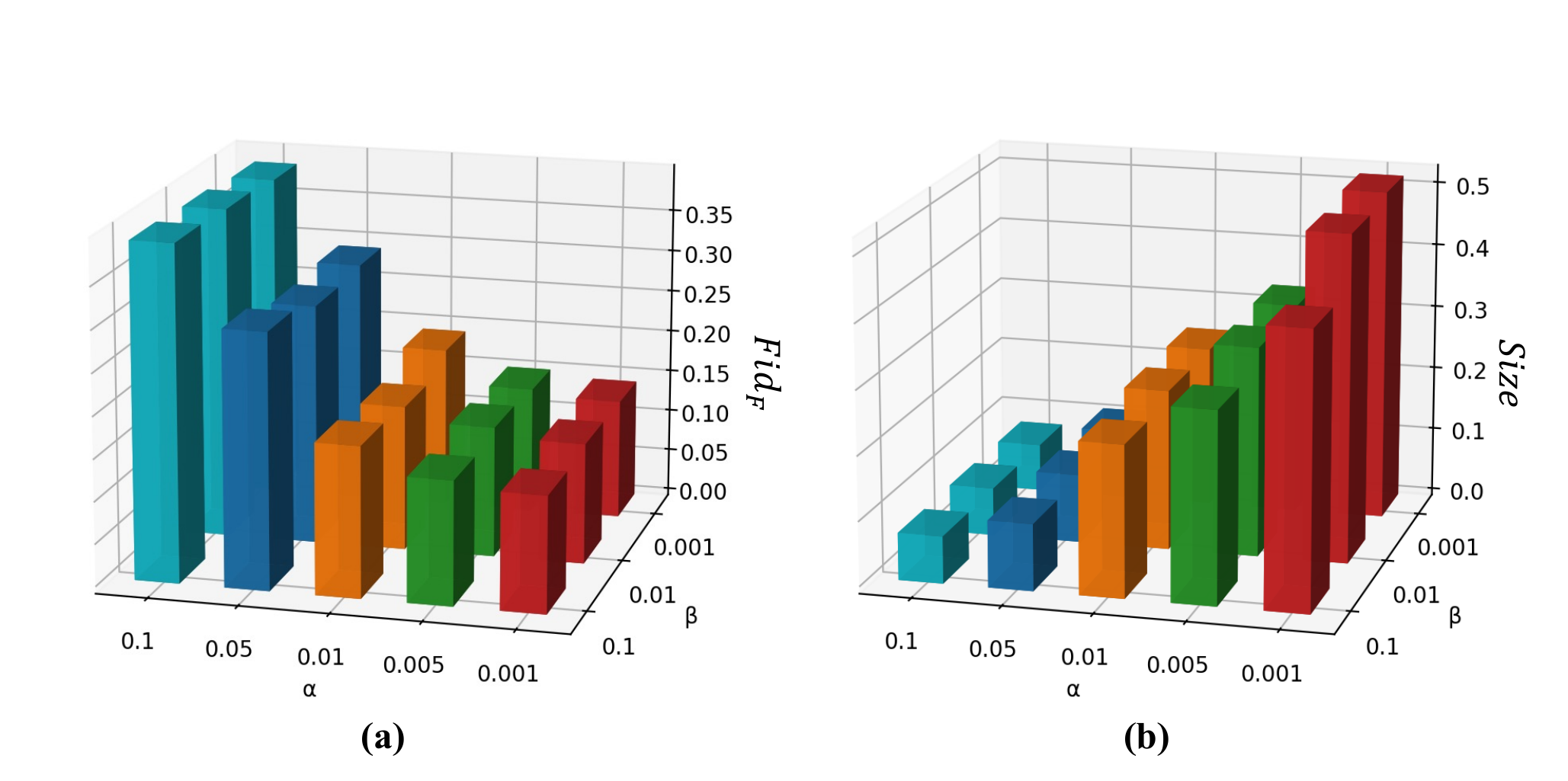}
    \caption{Sensitivity analysis of the hyperparameters $\alpha$ and $\beta$. (a) Effect on $Fid_{F}$, (b) Effect on $Size$.}
    \label{hyper}
\end{figure}

We conduct a hyperparameter sensitivity analysis using various combinations of hyperparameters on the validation dataset. The validation dataset is randomly sampled from the validation set of AudioCaps~\cite{kim2019audiocaps}. As illustrated in Figure~\ref{hyper}-(a), when the value of $\alpha$ decreases from 0.1 to 0.001, $Fid_{F}$ significantly decreases from 0.398 to 0.138. Furthermore, when $\beta$ is adjusted from 0.1 to 0.001 while keeping $\alpha = 0.001$, $Fid_{F}$ shows a slight increase from 0.138 to 0.142. This suggests that lowering $\beta$ can slightly enhance fidelity, but the effect is marginal compared to that of $\alpha$. In contrast, the impact on $Size$, shown in Figure~\ref{hyper}-(b), reveals a different trend. As $\alpha$ decreases from 0.1 to 0.001, $Size$ increases substantially from 0.07 to 0.51, indicating that lower $\alpha$ values lead to greater model complexity. Similarly, when $\beta$ is reduced from 0.1 to 0.001 with a fixed $\alpha = 0.001$, $Size$ further increases from 0.426 to 0.517. This demonstrates that both $\alpha$ and $\beta$ reductions tend to increase the mask size. These results indicate a trade-off between fidelity and size. To ensure fair comparisons with the baselines while maintaining a comparable mask size, we set $\alpha = 0.001$ and $\beta = 0.1$ based on this analysis.

\begin{figure*}
    \center
    \includegraphics[width=1.0 \linewidth]{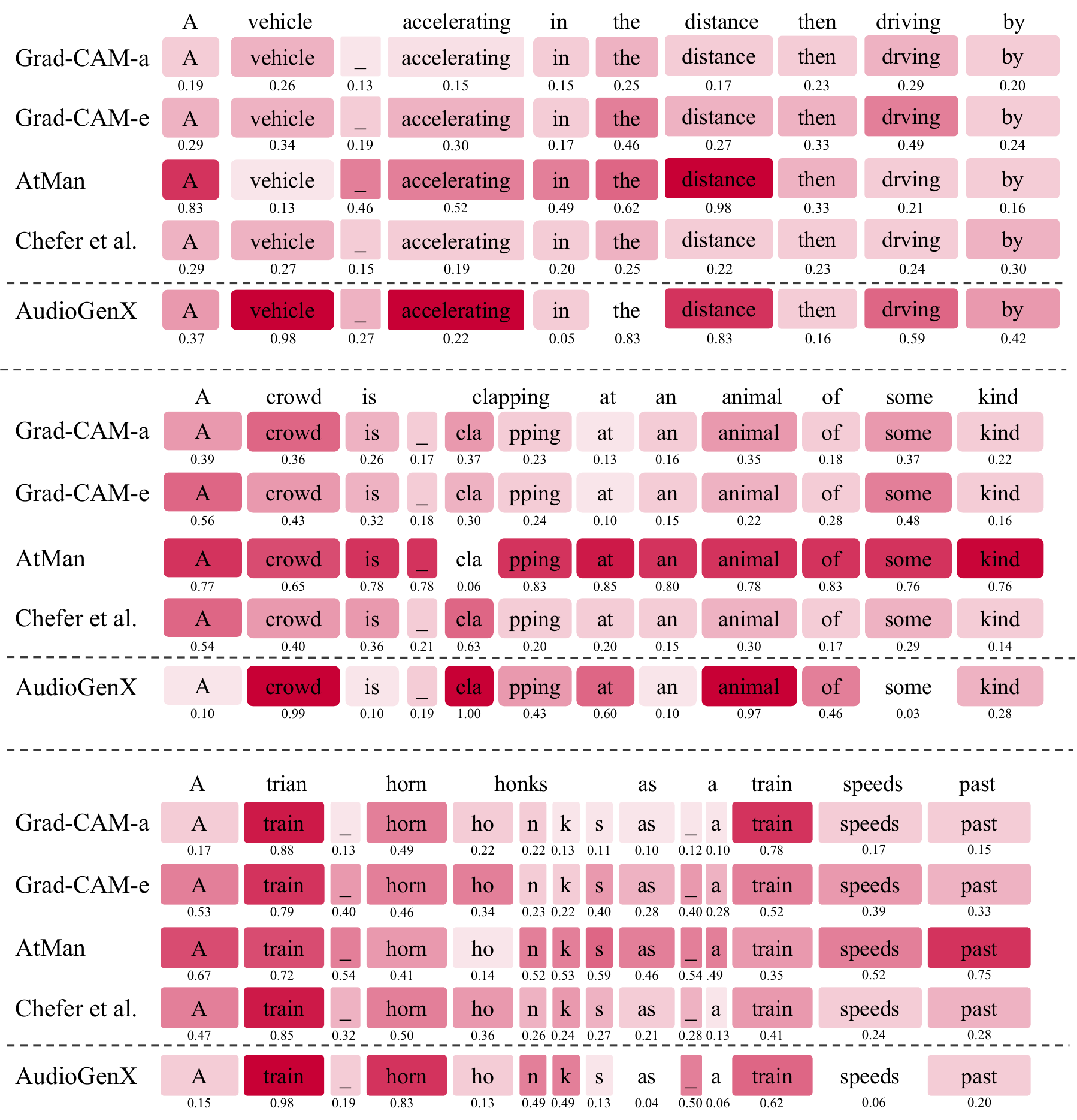}
    \caption{Qualitative analysis of \mname{} in comparison with baseline methods.}
    \label{experiment_appendix}
\end{figure*}

\begin{table*}[!ht]
    \centering
    \begin{tabular}{lllll|lllll}
    \toprule
      & \multicolumn{4}{c|}{Top 50 high-importance tokens}& \multicolumn{4}{c}{Top 50 low-importance tokens} \\
        Index & Textual token & Avg. Impt & Count & POS & Index & Textual token & Avg. Impt & Count & POS \\ 
        \midrule
        1 & sewing & 0.987 & 20 & VBG & 1 & ground & 0.037 & 11 & NN \\ 
        2 & horse & 0.976 & 13 & NN & 2 & series & 0.048 & 13 & NN \\ 
        3 & emergency & 0.957 & 13 & NN & 3 & electronic & 0.060 & 17 & JJ \\ 
        4 & ren & 0.946 & 18 & NNS & 4 & for & 0.086 & 12 & IN \\ 
        5 & baby & 0.939 & 22 & NN & 5 & before & 0.107 & 11 & IN \\ 
        6 & thunder & 0.936 & 18 & NN & 6 & background & 0.110 & 107 & NN \\ 
        7 & toilet & 0.936 & 13 & VBP & 7 & then & 0.116 & 173 & RB \\ 
        8 & soft & 0.932 & 12 & JJ & 8 & repeatedly & 0.119 & 12 & RB \\ 
        9 & rog & 0.931 & 13 & NN & 9 & while & 0.129 & 86 & IN \\ 
        10 & foot & 0.931 & 12 & NN & 10 & into & 0.130 & 42 & IN \\ 
        11 & app & 0.931 & 15 & NN & 11 & some & 0.139 & 45 & DT \\ 
        12 & food & 0.915 & 17 & NN & 12 & another & 0.141 & 19 & DT \\ 
        13 & step & 0.898 & 12 & NN & 13 & over & 0.146 & 19 & IN \\ 
        14 & infant & 0.895 & 13 & NN & 14 & distance & 0.163 & 49 & NN \\ 
        15 & ack & 0.893 & 13 & NN & 15 & with & 0.167 & 174 & IN \\ 
        16 & talking & 0.892 & 122 & VBG & 16 & the & 0.167 & 204 & DT \\ 
        17 & crying & 0.883 & 27 & VBG & 17 & followed & 0.185 & 277 & VBD \\ 
        18 & cla & 0.878 & 33 & NN & 18 & ongoing & 0.196 & 10 & VBG \\ 
        19 & pig & 0.875 & 13 & IN & 19 & loud & 0.224 & 53 & JJ \\ 
        20 & laughter & 0.873 & 18 & NN & 20 & are & 0.228 & 30 & VBP \\ 
        21 & goat & 0.873 & 12 & NN & 21 & and & 0.233 & 568 & CC \\ 
        22 & clo & 0.863 & 10 & NN & 22 & power & 0.250 & 12 & NN \\ 
        23 & pour & 0.863 & 10 & VBP & 23 & occurs & 0.260 & 15 & VBZ \\ 
        24 & duck & 0.859 & 12 & NN & 24 & sounds & 0.264 & 25 & NNS \\ 
        25 & door & 0.857 & 26 & NN & 25 & pitched & 0.264 & 12 & VBN \\ 
        26 & tapping & 0.848 & 11 & VBG & 26 & surface & 0.269 & 31 & NN \\ 
        27 & footsteps & 0.846 & 11 & NNS & 27 & several & 0.270 & 37 & JJ \\ 
        28 & bus & 0.845 & 11 & NN & 28 & from & 0.279 & 22 & IN \\ 
        29 & clicking & 0.842 & 14 & VBG & 29 & king & 0.287 & 62 & VBG \\ 
        30 & truck & 0.838 & 13 & NN & 30 & steam & 0.308 & 16 & JJ \\ 
        31 & scrap & 0.834 & 16 & JJ & 31 & sound & 0.309 & 18 & JJ \\ 
        32 & crowd & 0.823 & 31 & NN & 32 & through & 0.313 & 13 & IN \\ 
        33 & speaks & 0.817 & 97 & NNS & 33 & down & 0.325 & 11 & RP \\ 
        34 & boat & 0.817 & 12 & NN & 34 & two & 0.343 & 17 & CD \\ 
        35 & cat & 0.815 & 14 & NN & 35 & light & 0.351 & 13 & JJ \\ 
        36 & explosion & 0.813 & 12 & NN & 36 & ing & 0.352 & 342 & VBG \\ 
        37 & woman & 0.811 & 104 & NN & 37 & runs & 0.353 & 18 & NNS \\ 
        38 & music & 0.811 & 33 & NN & 38 & les & 0.353 & 16 & NNS \\ 
        39 & clan & 0.810 & 24 & NN & 39 & microphone & 0.355 & 48 & NN \\ 
        40 & speech & 0.810 & 44 & JJ & 40 & high & 0.359 & 20 & JJ \\ 
        41 & whistle & 0.803 & 14 & JJ & 41 & ting & 0.361 & 14 & VBG \\ 
        42 & water & 0.799 & 82 & NN & 42 & ses & 0.363 & 12 & VBZ \\ 
        43 & speaking & 0.794 & 163 & VBG & 43 & his & 0.364 & 35 & PRP\$ \\ 
        44 & men & 0.789 & 21 & NNS & 44 & ling & 0.364 & 160 & VBG \\ 
        45 & train & 0.785 & 35 & VBP & 45 & ving & 0.375 & 30 & VBG \\ 
        46 & rain & 0.782 & 29 & NN & 46 & metal & 0.377 & 45 & VBP \\ 
        47 & laughing & 0.778 & 40 & VBG & 47 & end & 0.382 & 1003 & NN \\ 
        48 & flush & 0.769 & 14 & NN & 48 & small & 0.383 & 12 & JJ \\ 
        49 & helicopter & 0.768 & 12 & NN & 49 & tires & 0.384 & 11 & NNS \\ 
        50 & talks & 0.766 & 31 & NNS & 50 & motor & 0.417 & 49 & NN \\         
        \bottomrule
    \end{tabular}
    \caption{Averaged importance (Impt) per textual token learned by \mname{}. The name of the POS (Part of Speech) is followed by the categories in NLTK. Count refers to the occurrence frequency in the test dataset of AudioCaps.}
    \label{tabler3}
\end{table*}

\begin{figure*}
    \center
    \includegraphics[width=1.0 \linewidth]{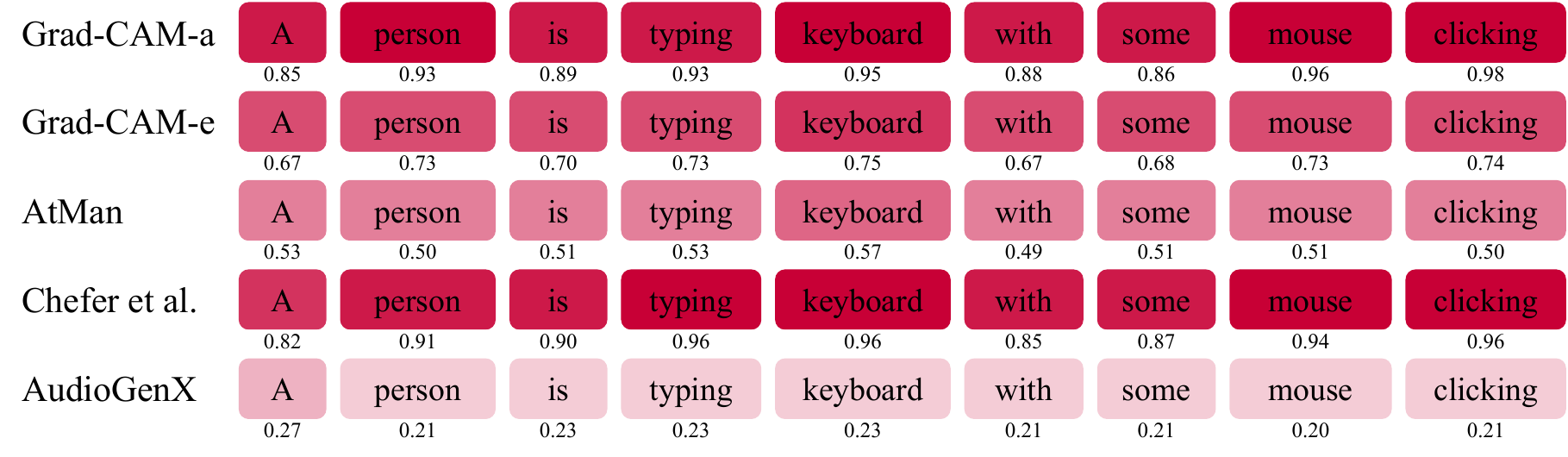}
    \caption{Examplar explanations using independent randomization on Transformer-decoder of AudioGen as the sanity check.}
    \label{sanitycheck}
\end{figure*}

\subsection{Leveraging Explanations to Understand Model Behavior and Edit Audio}
\label{casestudy}

\mname{} enhances transparency but also provides valuable insight for debugging TAG models and editing sound. In Table~\ref{tabler3}, we investigate the patterns of AudioGen using explanations generated by \mname{}. We first aggregated our explanations from the experiments in Table~\ref{tabler3}. Then, we filtered out tokens with a length greater than 1 and an occurrence frequency exceeding 10. From this subset, we selected the top 100 tokens with the highest average importance (Avg. Impt) and the bottom 100 tokens with the lowest average importance to generate word clouds. Detailed information on these tokens is presented in Table~\ref{tabler3}. The tokens in the top 100 predominantly consist of nouns (NN) and tokens that are associated with sounds. In contrast, the tokens in the bottom 100 displayed a diverse range of parts of speech, including adverbs (RB) and prepositions (IN), which tend to convey context rather than having intrinsic auditory significance. Additionally, as noted in the AudioGen documentation~\cite{kreuk2022audiogen}, tokens related to numbers (CD) or sequences also exhibit a lower importance.

Next, we utilize our explanations in the task of editing generated audio when it misaligns with the user's intended prompt. For example, given the prompt `Wind blows hard followed by screaming,' the generated audio should reflect this sequence. However, in our example, AudioGen fails to accurately capture the `screaming' sound. Using \mname{}, we find that `screaming' has low importance, especially in the latter part of the audio where it should be emphasized. To correct this, we used a technique in a study~\cite{hertz2022prompt} that adjusts attention weights to better align the audio with the intended prompt. The method of importance re-weighting is described as follows:

\begin{equation}
\textbf{M}^{*}_{\textbf{U}_{l}, \textbf{z}_{t}} := 
\begin{cases} 
c & \text{if } l=l^{*} \text{ and } t = t^{*}, \\ 
\textbf{M}_{\textbf{U}_{l}, \textbf{z}_{t}} & \text{otherwise},
\end{cases}
\end{equation}

\begin{table}[ht]
\centering
\begin{tabular}{l|l|l}
\toprule
   & FAD $\downarrow$ & $KL_{F} \downarrow$  \\ \hline
Before edit & 16.85 & 6.45 \\ 
After edit  & 2.68 & 1.82 \\ 
\bottomrule
\end{tabular}
\caption{Evaluation of editing generated audio}
\label{tab:edit}
\end{table}

\noindent where the explanation mask value $\textbf{M}_{\textbf{U}_{l}, \textbf{z}_{t}}$ from \mname{} is reweighted to $\textbf{M}^{*}_{\textbf{U}_{l}, \textbf{z}_{t}}$. In this case, $l^*$ and $t^*$ denote the target indices of the text token and the audio token, respectively.
When amplifying the explanation mask value, we set the scaling parameter $c$ to 0.9. Conversely, when suppressing the impact of the token, $c$ is set to 0.1. 
The threshold values (0.9 and 0.1) are chosen based on our heuristic intuition in Table~\ref{tabler3} that the importance of the top and bottom ranking is greater than 0.9 and less than 0.1, respectively.
For evaluation, we randomly sampled 100 prompts and identified 30 failure cases where AudioGen-generated audio differs from the ground truth in the AudioCaps~\cite{kim2019audiocaps} data set. To analyze these outputs, we applied \mname{} to identify which textual tokens were over- or underemphasized, then manually adjusted the importance of the tokens to align with the ground truth. 
As evaluation metrics, we compute the \text{Fr\'echet Audio Distance} (FAD)~\cite{fad} over both real and generated audio. FAD is an adaptation of the Fr\'echet Inception Distance (FID) for audio, measuring the similarity between distributions of real and generated audio data. Additionally, we measure the metric $KL_F$.

\begin{figure}
    \center
    \includegraphics[width=1.0 \linewidth]{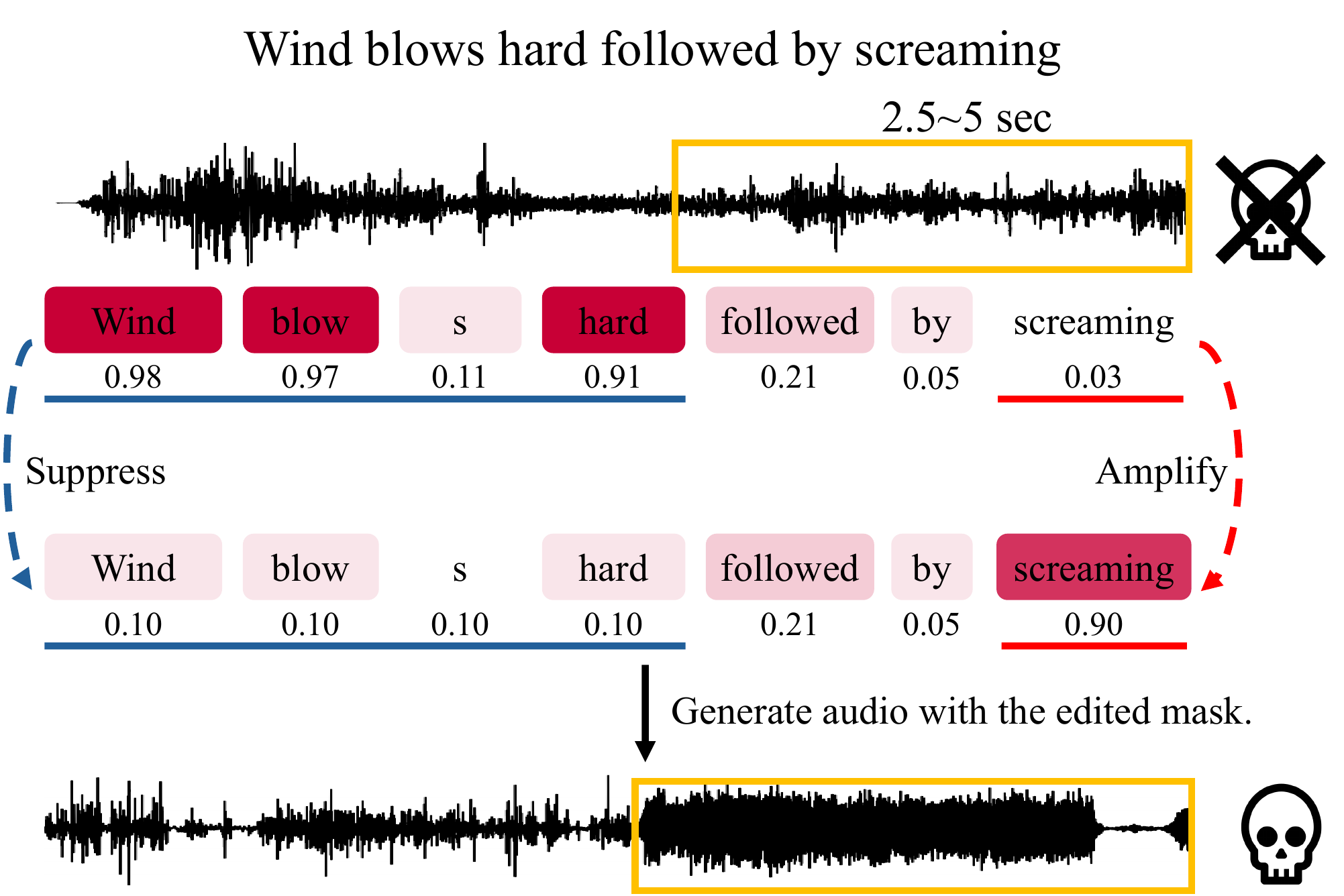}
    \caption{The scenario of editing the generated audio.}
    \label{fig:edit}
\end{figure}

Table~\ref{tab:edit} demonstrates that editing the importance mask allows us to generate audio that more closely matches the ground truth. For the setup, we randomly sampled 100 prompts, finding 30 failure cases that differed from the ground truth in the AudioCaps data set. Initially, the scores (FAD and KL) were lower than typical ones, at 3.13 and 2.09 in MAGNeT~\cite{ziv2024masked} known as the SOTA model. To understand the generated output, we used \mname{} to identify which textual tokens were over- or underemphasized, then manually adjusted the importance of the tokens to amplify or suppress. Figure~\ref{fig:edit} further illustrates that editing can be applied to specific time intervals. For instance, after re-weighting the mask values between 2.5 and 5 seconds, the ‘screaming’ audio emerges in the corresponding time interval. Although explanations do not directly involve generation, \mname{} helps users by offering valuable guidance during the editing process when there is a discrepancy between the user’s intention and the generated result.

\subsection{Sanity Check}
We conduct a sanity check following the approach in \cite{adebayo2018sanity} to assess the explanations generated. Specifically, we initialize the parameters of the Transformer-decoder, which predicts the next sequence of audio tokens. As shown in Figure~\ref{sanitycheck}, when the model is initialized randomly, the influence of each token in the visualization becomes nearly indistinguishable. This result of the explainer, including our baseline, in response to the state of the model parameters, suggests that \mname{} produces faithful explanations. Thus, we conclude that \mname{} generates reliable and trustworthy explanations, as validated by the sanity check.

\subsection{Limitation}
\label{app:limication}
While we introduce a novel approach to explaining generated audio in TAG models, there are some limitations to consider. First, \mname{} contains several hyperparameters that may require data set-specific tuning for optimal performance. Automating this process or reducing hyperparameter sensitivity would improve usability. Furthermore, biases present in the training data may be reflected in both the generated audio and the explanations. Without proper safeguards and responsible deployment practices, these biases could reinforce harmful stereotypes. As research into audio generation progresses, it is crucial to proactively develop robust bias detection methods and advocate for the ethical use of these powerful approaches. Despite these limitations and considerations, we believe that \mname{} represents a valuable step toward improving the interpretability and trustworthiness of TAG models.

\end{document}